\documentstyle[12pt]{article}
\newcommand{\pabar}{\not{\!\partial}}
\newcommand{\Od}{{\cal O}}

\newcommand{\tr}{\mbox{tr}}

\newcommand{\Dbar}{\not{\!{\!D}}}

\newcommand {\Dim}{\mbox{dim}}

\def\gappeq{\mathrel{\rlap {\raise.5ex\hbox{$>$}}
{\lower.5ex\hbox{$\sim$}}}}

\def\lappeq{\mathrel{\rlap{\raise.5ex\hbox{$<$}}
{\lower.5ex\hbox{$\sim$}}}}

\begin{document}
\input epsf \renewcommand{\topfraction}{0.8}
\pagestyle{empty}
\begin{flushright}
{CERN-TH/2000-195}
\end{flushright}
\vspace*{5mm}
\begin{center}
\Large{\bf The dynamics of the Goldstone bosons on the brane} \\
\vspace*{1cm}
\large{\bf Antonio Dobado}$^{1,*}$ {\bf and Antonio L. Maroto}
$^{1,2,\dagger}$ \\
\vspace{0.3cm}
\normalsize $^{1}$Departamento de  F\'{\i}sica Te\'orica,\\
 Universidad Complutense de
  Madrid, 28040 Madrid, Spain\\
\vspace{0.3cm}
$^{2}$CERN Theory Division, \\
CH-1211 Geneva 23, Switzerland\\
\vspace*{2cm}
{\bf ABSTRACT} \\ \end{center}
\vspace*{5mm}
\noindent

We study, within the recently proposed brane-world scenario, the
effective action for the low-energy brane excitations. These modes
are understood as Goldstone bosons associated to the spontaneous
{\it isometry} breaking, induced  on the bulk space by the
presence of the brane. Starting from the Nambu--Goto action for
the brane, we obtain a non-linear sigma model describing the
low-energy interactions of the Goldstone bosons and extend the
calculation up to  $\Od(p^4)$. We also discuss the Higgs-like
mechanism in which the Kaluza--Klein gauge fields absorb the
Goldstone bosons and acquire mass. Finally, we present the
explicit form of the effective action describing the low-energy
 interactions between the three-brane Goldstone bosons and the
particles present in the Standard Model. \vspace*{0.5cm}

\begin{flushleft} CERN-TH/2000-195 \\
July 2000
\end{flushleft}
\vspace*{0.5cm}

\noindent
\rule[.1in]{8cm}{.002in}

\noindent
$^{*}$E-mail: dobado@eucmax.sim.ucm.es\\
$^{\dagger}$E-mail: Antonio.Lopez.Maroto@cern.ch

\vfill\eject

\setcounter{page}{1}
\pagestyle{plain}

\section{Introduction}
The existence of large extra dimensions has been proposed
\cite{Hamed1} as a possible  explanation of the enormous
difference between the electroweak ($M_W$) and gravitational
($M_P$) scales, which typically is of the order of $M_P/M_W\simeq
10^{16}$. In this proposal, the most relevant scale would be the
Planck scale in $D=4+n$ dimensions $M_D$, which is assumed to be
not too far from the TeV scale in order to solve the
above-mentioned hierarchy problem. The much larger value of the
4-dimensional Planck constant would be due to the size of the
extra dimensions, since we typically have $M^2_P \simeq
R^nM_D^{n+2}$, where $R$ is the  radius of the compactified extra
dimensions. Phenomenological considerations then require that,
whereas gravity can propagate in the D-dimensional bulk space, the
ordinary matter fields and gauge bosons be bound to live on a
3-dimensional brane,  which would constitute the usual spatial
dimensions. The brane tension $f$ is the other important scale in
this scenario, $f^{-1}$ being the typical size of the brane
fluctuations.

A large number of works have been devoted to the study of the
phenomenological implications of this scenario (see \cite{Hamed2}
and references therein). In particular, special attention has been
paid to the description of the low-energy sector of the model.
This sector would include the Standard Model (SM) fields, the
gravitons and the possible excitations of the brane
\cite{Sundrum}. The presence of  extra dimensions allows for the
existence, in addition to the standard massless gravitons in four
dimensions, of an infinite tower of massive gravitons
(Kaluza--Klein \cite{KK} gravitons) whose mass is determined by
the size of the extra dimensions (see for example \cite{Balo} for
a review of different Kaluza--Klein models). The interaction of
the graviton sector with the SM fields has been analysed in a
series of papers \cite{Giudice} and different predictions have
been obtained that could be tested at future particle colliders.
However, in addition to the gravitons, the low-energy spectrum of
the theory also contains the brane's own excitations. If the brane
has been spontaneously created with a given shape in the bulk
(that we will consider as its ground state), the initial
isometries of the bulk space could be broken spontaneously by the
presence of the brane. The brane configurations obtained by means
of some isometry transformations in the bulk will be considered as
equivalent ground states and therefore the parameters describing
such transformations can be considered as zero-mode excitations of
the ground state. When such transformations are made local
(depending on the position on the brane), the corresponding
parameters play the role of Goldstone bosons (GB) fields of the
isometry breaking. Moreover, it has been shown that, in the case
where the brane tension $f$ is much smaller than the fundamental
scale $M_D$ $(f\ll M_D)$, the non-zero KK modes decouple from the
GB modes \cite{GB,Kugo} and it is then possible, at least in
principle, to make a low-energy effective theory description of
the GB dynamics. On the other hand, in the standard Kaluza--Klein
models, the isometries in the extra dimensions are understood as
gauge transformations in the four-dimensional theory. Therefore,
since the GB are associated to the breaking of those {\it gauge}
transformations, it is natural to expect \cite{Hamed2} that some
kind of Higgs mechanism can take place, which would  give mass to
the Kaluza--Klein gauge bosons.

In this paper we study the low-energy dynamics of these GB and the
Higgs-like mechanism we have just commented on. In section 2 we
set the main assumptions and the notation used in the rest of the
work. Starting from a Nambu--Goto like action for the brane, we
obtain the non-linear sigma model describing the low-energy
dynamics of the brane GB. Those GB correspond to the spontaneous
symmetry breaking of the translational isometries of  the
compactified extra dimensions by the brane.  When the isometries
are not exact but only approximate, the GB become pseudo-GB and
acquire some mass. Section 3 is devoted to the calculation of
these  masses. There we compute also the next to leading order
corrections to the non-linear sigma model containing four
derivatives of the GB fields.  In section 4 we describe in detail
how the GB can be absorbed by the KK graviphotons, giving rise to
the Higgs mechanism. In section 5 we study the couplings of GB
with the fields present in the SM, including scalars, (chiral)
fermions and gauge bosons. Finally, section 6 contains the main
conclusions of the work.

\section{The effective action for the Goldstone bosons}

In the following we will consider the four-dimensional space-time
$M_4$ to be embedded in a $D$-dimensional bulk space that for
simplicity we will assume to be of the form $M_D=M_4\times B$,
where $B$ is a given  compact manifold. This kind of spaces
include, as a particular case,  the extra-dimensional tori spaces
usually considered in the literature \cite{Kugo}. The coordinates
parametrizing the points in $M_D$ will be denoted by
$(x^{\mu},y^m)$, where the different indices run as $\mu=0,1,2,3$
and $m=4,5,...,D-1$. The bulk space $M_D$ is endowed with a metric
tensor that we will denote by $G_{MN}$, with signature
$(+,-,-...-,-)$. In the absence of the 3-brane, this metric
possesses an isometry group that we will assume to be of the form
$G(M_D)=G(M_4)\times G(B)$. The presence of the brane
spontaneously breaks this symmetry down to some subgroup
$G(M_4)\times H$. Therefore, we can introduce the coset space
$K=G(M_D)/(G(M_4)\times H) =G(B)/H$, where $H\subset G$ is a
certain subgroup of $G$ that we will study below.

Let us first consider the simplest case, in which the internal
manifold is just the circle $B=S^1$, i.e.  $M_5=M_4\times S^1$ and
the corresponding isometry group is just abelian $G(B)=U(1)$.  In
this case, the brane parametrization reads $Y^M=(x^{\mu},Y(x))$
and the metric tensor induced on the brane $g_{\mu\nu}$ is given
by:
\begin{equation}
g_{\mu\nu}=\partial_{\mu}Y^M\partial_{\nu}Y^NG_{MN}. \label{IM}
\end{equation}
We consider the simplest form for the metric on the bulk:
\begin{eqnarray*}
 G_{MN}&=&
\left(
\begin{array}{cccc}
\tilde g_{\mu\nu}(x)&0\\ 0&-1
\end{array}\right).
\end{eqnarray*}
According to (\ref{IM}), the induced metric on the brane is given
in this case by:
\begin{equation}
g_{\mu\nu}=\tilde g_{\mu\nu}-\partial_{\mu}Y\partial_{\nu}Y
\end{equation}
so that
\begin{equation}
\sqrt{g}=(-\det g)^{1/2}=\sqrt{\tilde g}\left(1-\frac{1}{2}\tilde
g^{\mu\nu}
\partial_{\mu}Y\partial_{\nu}Y+...\right).
\end{equation}
Under very general assumptions, the low-energy action for a narrow
3-brane can be taken as a Nambu--Goto action (originally
introduced in the context of membranes by Dirac \cite{Dirac}),
i.e.:
\begin{equation}
S_B=-f^4 \int_{M_4}d^4x\sqrt{g}, \label{Nambu}
\end{equation}
where $d^4x\sqrt{g}$ is the volume element of the brane and $f$ is
the brane tension. Thus for small excitations, the effective
action becomes
\begin{equation}
S_B=-f^4 \int_{M_4}d^4x\sqrt{\tilde g}
+\frac{f^4}{2}\int_{M_4}d^4x\sqrt{\tilde g} \tilde
g^{\mu\nu}\partial_{\mu}Y\partial_{\nu}Y.
\end{equation}
In this simple case there is only one Goldstone boson that
parametrizes the small excitations of the brane as $Y(x)$, and can
be identified with the coordinate in the internal dimension. As we
will see this is not always the case for more general  spaces $B$.

Let us thus consider internal spaces with higher dimensions. Then
the isometry group is in general non-abelian. We introduce the
$T_a$ ($a=1,...,\Dim(G(B))$) which are the generators of $G(B)$
with commutation relations:
\begin{equation}
[T_a,T_b]=iC_{abc}T_c,
\end{equation}
where $C_{abc}$ are the structure constants. The $B$ Killing
vectors $\xi_a(y)$ satisfy the $G(B)$ Lie algebra:
\begin{equation}
\{\xi_a,\xi_b\}=iC_{abc}\xi_c,
\end{equation}
where the brackets are the standard Lie brackets. The brane
parametrization in this case can be written as $Y^M=(x^\mu,
Y^m(x))$. The brane is created with a shape $Y^m(x)$ that
minimizes the action (\ref{Nambu}). In particular a possible
solution corresponds to $Y^m(x)=Y^m_0$, i.e. the brane is created
in a certain point in $B$. In this case the presence of the brane
will break spontaneously all the $B$ isometries, except for those
that leave the point $Y_0$ unchanged. In other words the group
$G(B)$ is spontaneously broken down to $H(Y_0)$, where $H(Y_0)$
denotes the isotropy group (or little group) of the point $Y_0$.
Let $H_i$ be the $H$ generators ($i=1,2,...\; h$), $X_\alpha$
($\alpha=1,2,...\;k=\mbox{dim}\; G-\mbox{dim}\;H$) the broken
generators, and $T=(H,X)$ the complete set of generators of $G$. A
similar separation can be done with the Killing fields. We will
denote $\xi_i$ those associated to the $H_i$ generators and
$\xi_\alpha$ those corresponding to $X_\alpha$. The excitations of
the brane along the (broken) Killing fields directions of $B$
correspond to the zero modes and they are parametrized by the GB
fields $\pi^\alpha(x)$ that can be understood as coordinates on
the coset manifold $K=G/H$. Thus, let us assume again that the
brane ground state is position-independent $Y^m_0$, then the
action of an element of $G(B)$ on $B$ will map $Y_0$ into some
other point with coordinates:
\begin{equation}
Y^m(x)=Y^m(Y_0,\pi^\alpha(x))=Y^m_0+\frac{1}{k
f^2}\xi^m_\alpha(Y_0)\pi^\alpha(x)+\Od(\pi^2)
\end{equation}
where we have set the appropriate normalization of the GB fields
and  Killing fields   with $k^2=16\pi /M_P^2$. It is important to
note that the coordinates of the transformed point depend only on
$\pi^\alpha(x)$, i.e. on the parameters of the transformations
corresponding to the broken generators. The rest of the
transformations (corresponding to the $H$ subgroup) leave the
vacuum unchanged and therefore they are not GB. Thus  not all the
isometries  will give rise to zero modes of the brane.

When the action of the isometry group $G(B)$ on $B$ is transitive,
i.e. when any pair of points in $B$ can be connected by an
isometry transformation, $B$ is said to be a homogeneous space. In
this case, the isotropy group of a given point is independent of
the particular point we have chosen, i.e. $H(Y_0)=H$, and, under
certain regularity conditions \cite{Helgason}, it is possible to
prove that $B$ is diffeomorphic to $K=G/H$. As a consequence, in
those cases, the number of GB coincides with the number of extra
dimensions ($\Dim \;B=\Dim\; K=\Dim\; G-\Dim\; H$). Typical
examples of this case are $B=T^n$, $S^n$ or $SU(N)$. However, if
$B$ is not a homogenous space, this result does not hold and the
number of GB (which equals $\Dim\; K$) will be in general smaller
than  $\Dim\; B$. Thus, for example, consider $B$ to be a
two-dimensional ellipsoid with axial symmetry. In this case,
$G(B)=U(1)$ and $\Dim\; G(B)=1$. The isotropy group now depends on
the particular point we choose. If $Y_0$ corresponds to any of the
two poles on the symmetry axis, then $H(Y_0)=U(1)$, i.e. $\Dim
\;H(Y_0)=1$ and the number of GB is zero. However, for a generic
point, $H(Y_0)=1$ or $H(Y_0)=Z_2$ for points on the equator,
therefore $\Dim \;H(Y_0)=0$ and the
 number of GB is one. In
any case $\Dim \; K\neq \Dim \; B$.

In order to find the GB effective action for arbitrary dimensions,
we consider the bulk metric ansatz as:
\begin{eqnarray*}
 G_{MN}&=&
\left(
\begin{array}{cccc}
\tilde g_{\mu\nu}(x)&0\\ 0&- g'_{mn}(y)
\end{array}\right).
\end{eqnarray*}
In the ground state, the induced metric on the brane is given by
the four-dimensional components of the bulk space metric, i.e.
$g_{\mu\nu}=\tilde g_{\mu\nu}=G_{\mu\nu}$. When brane excitations
are present, the induced metric is given by $g_{\mu\nu}=\tilde
g_{\mu\nu}-\partial_{\mu}Y^m\partial_{\nu}Y^ng'_{mn}$, where
\begin{equation}
\partial_{\mu}Y^m(x)=\frac{\partial Y^m}{\partial
\pi^\alpha}\partial_\mu \pi^\alpha=
\frac{1}{kf^2}\xi^m_\alpha(Y_0)\partial_{\mu}\pi^\alpha+\Od(\pi^2)
\end{equation}
and, accordingly:
\begin{equation}
g_{\mu\nu}=\tilde g_{\mu\nu}- g'_{mn}\frac{\partial
Y^m}{\partial\pi^\alpha}\frac{\partial
Y^n}{\partial\pi^\beta}\partial_{\mu}\pi^\alpha
\partial_{\nu}\pi^\beta.
\end{equation}
Introducing the tensor $h_{\alpha\beta}(\pi)$  as
\begin{equation}
h_{\alpha\beta}(\pi)=f^4 g'_{mn}(Y(\pi))\frac{\partial
Y^m}{\partial\pi^\alpha}\frac{\partial Y^n}{\partial\pi^\beta},
\end{equation}
we have
\begin{equation}
g_{\mu\nu}=\tilde
g_{\mu\nu}-\frac{1}{f^4}h_{\alpha\beta}(\pi)\partial_{\mu}\pi^\alpha
\partial_{\nu}\pi^\beta
\end{equation}
and, for the square root of the metric determinant, we get
\begin{equation}
\sqrt{g}=\sqrt{\tilde g}\left(1-\frac{1}{2f^4}\tilde
g^{\mu\nu}h_{\alpha\beta}(\pi)
\partial_{\mu}\pi^\alpha\partial_{\nu}\pi^\beta+...\right)
\label{det}
\end{equation}
so that the effective action for the Goldstone bosons up to
$\Od(p^2)$ is nothing but the non-linear sigma model corresponding
to a symmetry-breaking pattern $G\rightarrow H$:
\begin{equation}
S_B^{(2)}=-f^4 \int_{M_4}d^4x\sqrt{\tilde
g}+\frac{1}{2}\int_{M_4}d^4x\sqrt{\tilde g} \tilde
g^{\mu\nu}h_{\alpha\beta}(\pi)\partial_{\mu}\pi^\alpha\partial_{\nu}\pi^\beta.
\label{ea2}
\end{equation}

Notice that $\mu, \nu,...$  are $M_4$ indices, whereas
$\alpha,\beta,...$ are indices on the $K$ manifold. The above
expansion of the determinant in (\ref{det}) and therefore that of
the effective action in (\ref{ea2}) is not an expansion in powers
of the $\pi$ fields, but in powers of $\partial \pi/f^2$, i.e.
they are low-energy expansions.

\section{Mass terms and higher order corrections}

When the $G(M_D)$ symmetry is not exact, it can only be written as
the product $G(M_4)\times G(B)$ in an approximate way. In this
case, the Goldstone bosons are not massless, and their masses
measure  how accurately $G(M_D)$ describes the real symmetries of
the bulk space-time. This is similar to the $SU(2)_L\times
SU(2)_R$ chiral symmetries of the low-energy strong interactions.
This symmetry is spontaneously broken down to the isospin symmetry
$SU(2)_{L+R}$. The corresponding Goldstone bosons are the three
pions. However, the small quark masses break explicitly the chiral
symmetry and the above  breaking pattern is only approximate. The
pions then get a mass that is smaller than the rest of the hadron
masses in the QCD spectrum.

When the $G(M_D)$ symmetry is not exact, it is no longer possible
to assume that $\tilde g_{\mu\nu}$ depends only on $x$ and $g'_{m
n}$  on $y$. Thus, we can assume a slight dependence of $\tilde
g_{\mu\nu}$ also on the $y$ coordinates. We will use the following
ansatz in the Abelian case:
\begin{eqnarray*}
 G_{MN}&=&
\left(
\begin{array}{cccc}
\tilde g_{\mu\nu}(x,y)&0\\ 0&-1
\end{array}\right).
\end{eqnarray*}
It is now possible to expand $\tilde g_{\mu\nu}(x,y)$ around
$y=Y_0$, where we choose the coordinates so that $y=Y_0$
corresponds to the  appropriate vacuum when the $G(M_D)$ symmetry
is not exact:
\begin{eqnarray}
\tilde g_{\mu\nu}(x,Y)=\tilde g_{\mu\nu}(x,Y_0)+
\partial_Y\tilde g_{\mu\nu}(x,Y_0)(Y-Y_0)+\frac{1}{2}\partial_Y^2
\tilde g_{\mu\nu}(x,Y_0)(Y-Y_0)^2+...
\end{eqnarray}
Therefore, at the lowest non-trivial order we have
\begin{eqnarray}
\sqrt{g}=\sqrt{\tilde g}\left(1-\frac{1}{2}\tilde g^{\mu\nu}
\partial_{\mu}Y\partial_{\nu}Y+\frac{1}{4}\tilde g^{\mu\nu}
(\partial_Y^2\tilde g_{\mu\nu})Y^2+...\right).
\end{eqnarray}
Then, in terms of the properly normalized Goldstone boson field
$\pi=f^2 Y$, the low-energy effective action can be written as
\begin{eqnarray}
S_B =-f^4 \int_{M_4}d^4x\sqrt{\tilde g}
+\frac{1}{2}\int_{M_4}d^4x\sqrt{\tilde g} \tilde
g^{\mu\nu}\partial_{\mu}\pi\partial_{\nu}\pi
-\frac{1}{4}\int_{M_4}d^4x\sqrt{\tilde g}\tilde g^{\mu\nu}
(\partial_Y^2\tilde g_{\mu\nu})\pi^2+...
\end{eqnarray}
Thus the pseudo-Goldstone boson field $\pi$ gets a mass
\begin{equation}
M^2=\frac{1}{2}\tilde g^{\mu\nu} (\partial_Y^2\tilde g_{\mu\nu}).
\end{equation}
The general case can be studied in a similar way. We start from
the ansatz
\begin{eqnarray*}
 G_{MN}&=&
\left(
\begin{array}{cccc}
\tilde g_{\mu\nu}(x,y)&0\\ 0&- g'_{mn}(y)
\end{array}\right),
\end{eqnarray*}
where again $\tilde g_{\mu\nu}(x,Y_0)$ corresponds to the ground
state metric in the symmetric case. Expanding this metric around
$y^m=Y^m_0$ in terms of the $\pi^\alpha$ fields we find
\begin{eqnarray}
\tilde g_{\mu\nu}(x,Y)&=&\tilde g_{\mu\nu}(x,Y_0)+\partial_m
\tilde g_{\mu\nu}(x,Y_0)(Y^m-Y^m_0)\nonumber \\
&+&\frac{1}{2}\partial_m\partial_n\tilde
g_{\mu\nu}(x,Y_0)(Y^m-Y^m_0)(Y^n-Y^n_0)+...\nonumber \\ &=& \tilde
g_{\mu\nu}(x,Y_0) +\partial_m \tilde
g_{\mu\nu}(x,Y_0)\left(\frac{\xi^m_\alpha}{kf^2}\pi^\alpha
+\frac{1}{2}\left.\frac{\partial^2 Y^m}{\partial
\pi^\alpha\partial \pi^\beta}\right\vert_{\pi=0}
\pi^\alpha\pi^\beta\right)\nonumber \\&+&
\frac{1}{2}(\partial_m\partial_n\tilde
g_{\mu\nu}(x,Y_0))\frac{\xi^m_\alpha\xi^n_\beta}{k^2f^4}\pi^\alpha\pi^\beta
+\Od(\pi^3).
\end{eqnarray}
The effective action is then given by
\begin{eqnarray}
S_B &=&-f^4 \int_{M_4}d^4x\sqrt{\tilde g}  \nonumber \\
&+&\frac{1}{2}\int_{M_4}d^4x\sqrt{\tilde g} (\tilde
g^{\mu\nu}h_{\alpha\beta}(\pi)\partial_{\mu}\pi^\alpha\partial_{\nu}\pi^\beta
-M^2_{\alpha\beta} \pi^\alpha\pi^\beta)+...,
\end{eqnarray}
where $\tilde g^{\mu\nu}$ denotes $\tilde g^{\mu\nu}(x,Y_0)$ and
the mass matrix can be written as
\begin{equation}
M^2_{\alpha\beta}=\frac{f^4}{2}\tilde
g^{\mu\nu}\left(\left.\frac{\partial^2 Y^m}{\partial
\pi^\alpha\partial \pi^\beta}\right\vert_{\pi=0}\partial_m \tilde
g_{\mu\nu}+\partial_m\partial_n\tilde
g_{\mu\nu}\frac{\xi^m_\alpha\xi^n_\beta}{k^2f^4}\right).
\end{equation}

From the above  discussion  we arrive at the conclusion that the
pseudo-GB masses are of the order of magnitude of the derivatives
of the physical four-dimensional metric, which are expected to be
of the order of the inverse brane size fluctuations $f^{-1}$, i.e.
we expect the pseudo-GB masses to be of the same order of
magnitude as the brane tension $f$.

Finally we will extract the $\Od(p^4)$ corrections to the GB
effective
 action.
With that purpose we further expand the metric determinant
$\sqrt{g}$ in terms of the metric $\tilde g_{\mu\nu}$ and of the
GB modes derivatives $\partial \pi^\alpha$, following  steps
similar to what was done in (\ref{det}). Thus we find:
\begin{eqnarray}
\sqrt g &=& \sqrt{ \tilde g}(1-\frac{1}{2f^4} \tilde
g^{\mu\nu}h_{\alpha\beta}(\pi)\partial_{\mu}\pi^\alpha
\partial_{\nu}\pi^\beta    \nonumber \\
&+& \frac{1}{8f^4}\tilde g^{\mu\nu}\tilde
g^{\rho\sigma}h_{\alpha\beta}(\pi)h_{\gamma\delta}(\pi)
\partial_{\mu}\pi^\alpha\partial_{\nu}\pi^\beta
\partial_{\rho}\pi^\gamma\partial_{\sigma}\pi^\delta \nonumber \\
&-& \frac{1}{4f^4}\tilde g^{\mu\nu}\tilde
g^{\rho\sigma}h_{\alpha\beta}(\pi)h_{\gamma\delta}(\pi)
\partial_{\nu}\pi^\alpha\partial_{\rho}\pi^\beta
\partial_{\sigma}\pi^\gamma\partial_{\mu}\pi^\delta+...),
\end{eqnarray}
so that the $\Od(p^4)$ effective action becomes
\begin{eqnarray}
S_B^{(4)} &=& -f^4 \int_{M_4}d^4x\sqrt{\tilde g}
+\frac{1}{2}\int_{M_4}d^4x\sqrt{\tilde g} \tilde
g^{\mu\nu}h_{\alpha\beta}(\pi)\partial_{\mu}\pi^\alpha\partial_{\nu}
\pi^\beta \nonumber
\\ &-&\frac{1}{8f^4}\int_{M_4}d^4x\sqrt{\tilde g} (\tilde
g^{\mu\nu}h_{\alpha\beta}(\pi)\partial_{\mu}\pi^\alpha\partial_{\nu}\pi^\beta)^2\nonumber
\\ &+&\frac{1}{4f^4}\int_{M_4}d^4x\sqrt{\tilde g} \tilde
g^{\mu\nu}\tilde
g^{\rho\sigma}h_{\alpha\beta}(\pi)h_{\gamma\delta}(\pi)
\partial_{\nu}\pi^\alpha\partial_{\rho}\pi^\beta
\partial_{\sigma}\pi^\gamma\partial_{\mu}\pi^\delta.
\end{eqnarray}
The GB self-interactions are organized in increasing number of GB
field derivatives over $f$. We thus have a well-defined energy
expansion so that, at low energies  with respect to $f$, only the
first terms contribute, in much the same way as it happens in the
Chiral Perturbation Theory ($\chi$PT) which is used to describe
the low-energy interactions of pions or the Electroweak Chiral
Lagrangians used to describe the symmetry breaking sector of the
SM (see for instance \cite{Dobado} and references therein). This
kind  of description is completely determined (at the lowest
order) by just the symmetry-breaking pattern $K=G/H$ and an energy
scale, $f$ in our case. However, in those examples of non-linear
sigma model effective theories, the coefficients of the
higher-derivative terms are undetermined and they have to be
obtained phenomenologically since the underlying theory cannot be
solved (QCD) or is unknown (symmetry-breaking sector). However, in
the case considered here, these coefficients are explicitly
obtained  in terms of the brane tension $f$ from the underlying
theory, which is given by the Nambu--Goto-like brane action.

\section{Kaluza--Klein gauge bosons and the Higgs mechanism}

As commented in the introduction, the isometries in the $B$ space
are considered as gauge transformation in the Kaluza--Klein
theories \cite{Balo}. In this section we study under which
circumstances the GB associated to the isometry breaking can give
rise to the longitudinal components of the Kaluza--Klein gauge
bosons, as in the standard Higgs mechanism.

 We start with the Hilbert--Einstein action for the gravitational field
 in $D$ dimensions  plus the brane action: $S=S_G+S_B$, i.e.
\begin{equation}
S=\frac{-1}{16\pi G_D}\int_{M_D} d^Dz\sqrt{G} R_D
-\frac{f^4}{4} \int_{M_4}d^4x\sqrt{g}
G_{MN}g^{\mu\nu}\partial_{\mu}Y^M\partial_{\nu}Y^N,
\end{equation}
where $z=(x,y)$ are the coordinates defined on $M_D$, $x$ and $y$
being the coordinates defined on $M_4$ and $B$ respectively, and
$R_D$ is the $D$-dimensional scalar curvature. For the particular
case of an $S^1$ compactified extra dimension (Abelian case),
$M_D=M_4 \times S^1$ and $G(B)=U(1)$. As usually done in the
Kaluza--Klein approach, we consider the metric ansatz
\begin{eqnarray*}
 G_{MN}&=&
\left(
\begin{array}{cccc}
\tilde g_{\mu\nu}(x)-B_{\mu}(x)B_{\nu}(x)&B_{\mu}(x)\\
B_{\nu}(x)&-1
\end{array}\right).
\end{eqnarray*}
According to the Kaluza--Klein philosophy, the translations on the
compactified fifth dimension $y\rightarrow y'=y + R \theta(x)$,
where $R$ is the compactification radius, can be understood as
standard gauge $U(1)$ transformations. With the above metric, the
five-dimensional gravitational action contains the terms
\begin{equation}
S_G=\frac{-1}{16\pi G_N}\int d^4x\sqrt{\tilde g} \tilde
R-\frac{1}{4} \int d^4x \sqrt{\tilde g} F_{\mu\nu}F^{\mu\nu}
\end{equation}
where $\tilde R$ is the scalar curvature corresponding to the
$\tilde g_{\mu\nu}$ metric, $G_5$ and $G_N=1/M_P^2$ are related by
$G_5= 2\pi R G_N$,  $F_{\mu\nu}=
\partial_{\mu}A_{\nu}-\partial_{\nu}A_{\mu}$ and $B_{\mu}=k A_{\mu}$.
On the other hand, the induced metric $g_{\mu\nu}$ turns out, in
this case, to be
\begin{equation}
g_{\mu\nu}=\tilde g_{\mu\nu}-D_{\mu}YD_{\nu}Y
\end{equation}
and
\begin{eqnarray}
\sqrt{g}=\sqrt{\tilde g}\left(1-\frac{1}{2}\tilde g^{\mu\nu}
D_{\mu}YD_{\nu}Y+...\right),
\end{eqnarray}
where the covariant derivative is defined as
\begin{equation}
D_{\mu}Y=\partial_{\mu}Y-B_{\mu}=\frac{1}{f^2}(\partial_{\mu}\pi-kf^2A_{\mu})
=\frac{1}{f^2}D_{\mu}\pi.
\end{equation}
Thus the total low-energy action can be written as
\begin{eqnarray}
S &=& \frac{-1}{16\pi G_N}\int d^4x\sqrt{\tilde g} \tilde
R-\frac{1}{4} \int d^4x \sqrt{\tilde g} F_{\mu\nu}F^{\mu\nu}
\nonumber \\ &-&f^4 \int_{M_4}d^4x\sqrt{\tilde
g}+\frac{f^4}{2}\int_{M_4}d^4x\sqrt{\tilde g} \tilde
g^{\mu\nu}D_{\mu}YD_{\nu}Y.
\end{eqnarray}
As a consequence, the gauge bosons (graviphotons) get a mass term
\begin{equation}
\frac{k^2f^4}{2}\int d^4x\sqrt{\tilde g}A_{\mu}A^{\mu},
\end{equation}
i.e. the gauge boson mass is $M^2=k^2f^4=16\pi\frac{f^4}{M_P^2}$,
which is typically very small. For example, in order to be able to
generate the $W$ mass through this Higgs mechanism, the brane
tension should be as large as the geometrical average between the
electroweak and the Planck scale. More precisely
\begin{equation}
f^2=\frac {M_WM_P}{4 \sqrt{\pi}}.
\end{equation}
In the general non-Abelian case, we consider again the brane and
space-time structure $M_D=M_4\times B$. As usual in the
Kaluza--Klein approach the metric ansatz is taken to be:
\begin{eqnarray*}
 G_{MN}&=&
\left(
\begin{array}{cccc}
\tilde
g_{\mu\nu}(x)-g'_{mn}(y)B_{\mu}^m(x,y)B_{\nu}^n(x,y)&B_{\mu}^n(x,y)\\
B_{\nu}^m(x,y)&-g'_{mn}(y)
\end{array}\right),
\end{eqnarray*}
where $B_{\mu}^n(x,y)=\xi^n_a(y)A_{\mu}^a(x)$ with $\xi^n_a(y)$
the Killing vectors
 corresponding to the isometry group $G(B)$ introduced above. Now the
gauge transformations are $y^m \rightarrow
y'^m=y^m+\xi^m_a(y)\epsilon^a(x)$. As is well known in this case,
the gravitational action $S_G$ can be written as
\begin{equation}
S_G=\frac{-1}{16\pi G_N}\int d^4x\sqrt{\tilde g} \tilde R-
\frac{\langle \xi^n_a\xi^m_b g'_{mn}\rangle }{16\pi G_N}
\frac{1}{4} \int_{M_4}d^4x \sqrt{\tilde g} F^a_{\mu\nu}F^{\mu\nu
b},
\end{equation}
where
$F^a_{\mu\nu}=\partial_{\mu}A^a_{\nu}-\partial_{\nu}A^a_{\mu}-
C_{abc}A^b_{\mu} A^c_{\nu}$ and
\begin{equation}
16 \pi G_N=\frac{16 \pi G_D}{\int_{B}d^{D-4} y  \sqrt{g'}}.
\end{equation}
In order to obtain the standard Yang--Mills action, the
normalization of the Killing vectors is given by
\begin{equation}
\langle\xi_a^m(y)\xi_b^n(y)g'_{mn}(y)\rangle=k^2\delta_{ab},
\label{norm}
\end{equation}
where $g'_{mn}(y)$ is the $B$ space-time metric, again $k^2=16\pi
G_N$ and the brackets are defined as the $B$ manifold average
\cite{Weinberg}
\begin{equation}
\langle f(y) \rangle= \frac{ \int_{B}d^{D-4} y\sqrt{g'}f(y) }{
\int_{B}d^{D-4}y\sqrt{g'} }.
\end{equation}

 The induced metric is
\begin{equation}
g_{\mu\nu}=\tilde g_{\mu\nu}-\Delta_{\mu}Y^m\Delta_{\nu}Y^ng'_{mn}
\end{equation}
and
\begin{eqnarray}
\sqrt{g}=\sqrt{\tilde g}\left(1-\frac{1}{2}\tilde g^{\mu\nu}
\Delta_{\mu}Y^m\Delta_{\nu}Y^n g'_{mn}+...\right),
\end{eqnarray}
where the covariant derivative is defined as
\begin{equation}
\Delta_{\mu}Y^m=\partial_{\mu}Y^m-\xi^m_aA_{\mu}^a,
\end{equation}
which can be written as
\begin{equation}
\Delta_{\mu}Y^m=\frac{\partial Y^m}{\partial \pi^\alpha}
\partial_{\mu}\pi^\alpha-\xi^m_a A_{\mu}^a=
\frac{1}{kf^2}\xi^m_\alpha(Y_0) (\partial_\mu \pi^\alpha-kf^2
A_\mu^\alpha)-\xi^m_i(Y_0) A_\mu^i+\Od(\pi^2).
\end{equation}
Since the $i$ indices correspond to the generators of the isotropy
group $H(Y_0)$, the Killing fields vanish at $Y_0$, i.e.
$\xi_i^m(Y_0)=0$ and the last term vanishes.

Therefore the brane action $S_B$ is
\begin{equation}
S_B=-f^4 \int_{M_4}d^4x\sqrt{\tilde g}
+\frac{1}{2}\int_{M_4}d^4x\sqrt{\tilde g} \tilde
g^{\mu\nu}h_{\alpha\beta}D_{\mu}\pi^\alpha
D_{\nu}\pi^\beta+\Od(\pi^4).
\end{equation}
where $D_\mu\pi^\alpha=\partial_\mu \pi^\alpha-kf^2 A_\mu^\alpha$.
Thus the gauge boson mass matrix is
\begin{equation}
M_{\alpha\beta}^2=k^2f^4h_{\alpha\beta}(0).
\end{equation}
Remember that $Y^m(x)=Y^m_0$ corresponds to $\pi^\alpha=0$. As
commented on before, not all the the gauge bosons will acquire a
mass by this mechanism. Only those associated to the broken
$X_\alpha$ generators will, their number being determined by the
dimension of the $K=G/H$ space. In any case, two important
comments are in order. First, as it happened in the abelian case,
the gauge boson masses are quite small whenever $f \ll M_P$. On
the other hand it should be remembered that in the standard KK
picture, having gauge couplings $g$ small enough to have a
sensible weak coupling limit (say $g<1$) requires having extra
dimensions  of a typical size of the order of the Planck length,
since $g^2$ is of the order of $k^2/R^2$ (see for instance
\cite{Balo}). Thus, for the interesting case of large extra
dimensions and $f \ll M_D$, graviphotons can be considered
massless and decoupled from the rest of the low-energy particles.
In this case we can safely assume that the Higgs mechanism has not
taken place and the GB can be considered as the only relevant
low-energy new degrees of freedom.

\section{Couplings to the Standard Model fields}
As we have shown in the previous sections, the induced metric on
the brane depends on both the four-dimensional bulk metric
components $\tilde g_{\mu\nu}$ and the Goldstone bosons
$\pi^\alpha$. In the following, we will assume that the physical
space-time metric is $\tilde g_{\mu\nu}$,  whereas the
contribution of the GB to the brane metric can be detected only
through their couplings to the Standard Model fields. In order to
obtain such couplings, we start from the Sundrum effective action
for the SM fields \cite{Sundrum}, which is basically the SM action
defined on a curved space-time $M_4$ whose metric is the induced
metric $g_{\mu\nu}$. Let us give the results for the different
kinds of fields (we will follow the notation in \cite{Dobado}):

\subsection*{Scalars}

 For a scalar field we start from the action
\begin{equation}
S_{\Phi}=\frac{1}{2}\int_{M_4}d^4x\sqrt{
g}g^{\mu\nu}\partial_{\mu}\Phi
\partial_{\nu}\Phi,
\end{equation}
which can be written  up to $\Od(p^2)$ as
\begin{eqnarray}
S_{\Phi}^{(2)} &=&\frac{1}{2}\int_{M_4}d^4x\sqrt{ \tilde g}\tilde
g^{\mu\nu}
\partial_{\mu}\Phi\partial_{\nu}\Phi    \nonumber \\
&+& \frac{1}{2f^4}\int_{M_4}d^4x\sqrt{ \tilde
g}h_{\alpha\beta}(\pi)
\partial_{\mu}\Phi\partial_{\nu}\Phi\partial^{\mu}\pi^\alpha
\partial^{\nu}\pi^\beta
 \nonumber \\
&-& \frac{1}{4f^4}\int_{M_4}d^4x\sqrt{ \tilde g} \tilde
g^{\mu\nu}\partial_{\mu}\Phi\partial_{\nu}\Phi \tilde
g^{\rho\sigma} h_{\alpha\beta}(\pi)
\partial_{\rho}\pi^\alpha \partial_{\sigma}\pi^\beta.
\end{eqnarray}

\subsection*{Fermions}

In order to introduce fermions on the brane, we will first extend
the vielbein formalism to $M_4\times B$. For this purpose we give
the necessary notation for the different kinds of indices. We will
use $A,B,... =0,..., D-1$ to denote flat indices in the bulk,
$M,N,...=0,..., D-1$ for curved indices in the bulk,
$i,j,...=0,...,3$ flat indices in $M_4$ (not to be confused with
the gauge indices for $H$ introduced in Sect. 2),
$\mu,\nu,...=0,...,3$ curved indices in $M_4$, $\bar m, \bar
n,...=4,..., D-1$ flat indices in $B$ and $m,n,...=4,..., D-1$
curved indices in $B$. Let us consider the following ansatz for
the vielbein in $M_4 \times B$:
\begin{eqnarray*}
 E^A_{\;M}&=&
\left(
\begin{array}{cccc}
\tilde e^i_{\;\mu}& B^{\bar m}_{\; \mu}\\ 0& e^{\bar m}_{\; m}
\end{array}\right)
\end{eqnarray*}
and the inverse vielbein:
\begin{eqnarray*}
 E_A^{\;M}&=&
\left(
\begin{array}{cccc}
\tilde e_i^{\;\mu}& -B_i^{\; m}\\ 0& e_{\bar m}^{\; m}
\end{array}\right).
\end{eqnarray*}
With these definitions we have: $G_{MN}=\eta_{AB} E^A_{\;
M}E^B_{\; N}$, $\tilde g_{\mu\nu}=\eta_{ij}\tilde
e^i_{\;\mu}\tilde e^j_{\;\nu}$ and $g_{mn}=\eta_{\bar m \bar
n}e^{\bar m}_{\; m}e^{\bar n}_{\; n}$, where $g_{mn}=-g'_{mn}$ and
$\eta_{\bar m \bar n}=-\delta_{\bar m \bar n}$.

In order to obtain the induced vierbein on the brane
$e^i_{\;\mu}$, we follow Sundrum \cite{Sundrum} in defining:
\begin{eqnarray}
e^i_{\;\mu}=R^i_{\; A}E^A_{\; M}\partial_\mu Y^M, \label{IV}
\label{viel}
\end{eqnarray}
where $R^A_{\; B}$ are the components of a Lorentz transformation
in the D-dimensional tangent space given by:
\begin{eqnarray}
R(x)=\exp (i\theta_{i \bar m}(x)J^{(i \bar m)}), \label{LT}
\end{eqnarray}
with $\theta_{i \bar m}(x)$ the transformation parameters and
$J^{(AB)}$ the generators of Lorentz transformations in $D$
dimensions. In particular, they can be written as:
\begin{eqnarray}
J^{(AB)C}_{\;\;\;D}=i\eta^{CE}(\delta^A_{\;E}
\delta^B_{\;D}-\delta^A_{\;D}\delta^B_{\;E}).
\end{eqnarray}
We will only consider those transformations $J^{(i \bar m)}$
mixing internal and external indices. Let us take those tangent
vectors given by $E^A_{\; M}(Y)\partial_\mu Y^M$, and act on them
with the Lorentz transformation given in (\ref{LT}). If we impose
that the transformed vectors be orthogonal to the $\bar m$
directions, i.e. they only have non-vanishing  four-dimensional
$i$ components, they will then satisfy:
\begin{eqnarray}
R^{\bar m}_{\; A}E^{A}_{\; M}(Y) \partial_\mu Y^M=0 \label{CO}
\end{eqnarray}
If $R^{\bar m}_{\; A}$ satisfies this condition, the induced
vierbein defined in (\ref{IV}) then possesses the properties of a
vierbein, i.e. $e^i_{\;\mu}e^j_{\;\nu}\eta_{ij}=g_{\mu\nu}$, where
$g_{\mu\nu}$ is the induced metric on the brane given in
(\ref{IM}). This definition allows us to introduce chiral fermions
in the brane in a straightforward way, as we will show below.

Using the expression for the vielbein given in (\ref{viel}), we
can find:
\begin{eqnarray}
e^i_{\;\mu}=R^i_{\;j}\tilde e^j_{\;\mu}+ R^i_{\;\bar p} e^{\bar
p}_{\;m}(\partial_\mu Y^m+B^m_{\; \mu}).
\end{eqnarray}
The expressions for  $R^i_{\;A}$ are determined from eqs.
(\ref{CO}). In particular, since we are mainly interested in the
couplings of the Goldstone bosons to fermions to lowest order, it
will be sufficient to calculate the components $R^i_{\;A}$ to
first (second)
 order in $\theta_{i \bar m}(x)$. Thus, in particular, we
will need:
\begin{eqnarray}
R^i_{\; j}&=&\delta^i_{\;j}-\frac{1}{2}\theta_{k \bar p}\theta_{j
\bar q}\eta^{\bar p \bar q}\eta^{ik}\nonumber \\ R^i_{\;\bar
p}&=&-\theta_{j \bar p}\eta^{ij}\nonumber \\R^{\bar
q}_{\;i}&=&\theta_{i \bar p}\eta^{\bar p \bar q} \nonumber \\
R^{\bar q}_{\; \bar m}&=&\delta^{\bar q}_{\;\bar m
}-\frac{1}{2}\theta_{j \bar p}\theta_{k \bar m}\eta^{\bar p \bar
q}\eta^{jk}.
\end{eqnarray}
From (\ref{CO}) we get:
\begin{eqnarray}
\theta_{i \bar q}=-\tilde e_i^{\;\mu}(B_{\bar q \mu}+\eta_{\bar q
\bar m}e^{\bar m}_{\; m}\partial_\mu Y^m).
\end{eqnarray}
Using this result and the definition of the induced vierbein in
(\ref{IV}), we obtain, up to second order in the gauge fields and
the derivatives of the coordinates:
\begin{eqnarray}
e^i_{\;\mu}&=&\tilde e^i_{\mu}+\frac{1}{2}T^i_{\;\mu}\nonumber \\
e_i^{\;\mu}&=&\tilde e_i^{\;\mu}-\frac{1}{2}T_i^{\;\mu},
\end{eqnarray}
where
\begin{eqnarray}
T_{\mu\nu}&=&-g'_{mn}(\partial_\mu Y^m+B_\mu^{\;m})(\partial_\nu
Y^n+B_\nu^{\;n})\nonumber \\ T^i_{\mu}&=&\tilde e^i_{\;\nu}\tilde
g^{\nu\rho}T_{\mu\rho}; \label{TT}
\end{eqnarray}
with these definitions the induced metric can be written as
$g_{\mu\nu}=\tilde g_{\mu\nu}+T_{\mu\nu}$.

We also need to know the expression of the induced spin
connection, defined as:
\begin{eqnarray}
\omega^{i\;\;j}_{\;\mu}=e^{i}_{\;\nu}\eta^{jk}(\partial_\mu
e_{k}^{\;\nu}+e_{k}^{\;\lambda}\Gamma^\nu_{\mu\lambda}),
\end{eqnarray}
where the induced Christoffel symbols on the brane are given by:
\begin{eqnarray}
\Gamma^\nu_{\mu\lambda}=\tilde\Gamma^\nu_{\mu\lambda}-T^{\nu\rho}\tilde
g_{\rho\sigma}\tilde\Gamma^\sigma_{\mu\lambda}+\frac{1}{2}\tilde
g^{\nu\rho}(\partial_\mu T_{\rho\lambda}+\partial_\lambda
T_{\rho\mu}-\partial_\rho T_{\mu\lambda}).
\end{eqnarray}
With these expressions we can obtain the form of the induced spin
connection in terms of the spin connection asscociated to the
metric $\tilde g_{\mu\nu}$. We have, up to first order in $T$:
\begin{eqnarray}
\omega^{i\;\;j}_{\;\mu}=\tilde
\omega^{i\;\;j}_{\;\mu}+\frac{1}{2}\tilde e^{i}_{\;\nu}\tilde
e_{k}^{\;\lambda} \eta^{jk}\tilde g^{\nu\rho}(T_{\rho\mu ;
\lambda}-T_{\mu\lambda ; \rho})
\end{eqnarray}
The Dirac action for a massless fermion on the brane can be
written as:
\begin{eqnarray}
S=\int d^4 x \sqrt{g} i\bar \psi
e_i^{\;\mu}\gamma^i(\partial_\mu+\Omega_\mu)\psi,
\end{eqnarray}
where
$\Omega_\mu=\frac{1}{8}\omega^{i\;\;j}_{\;\mu}[\gamma_i,\gamma_j]$
and $\{\gamma^i,\gamma^j\}=2\eta^{ij}$. Using the above equations
we can expand the Dirac action in terms of the metric $\tilde
g_{\mu\nu}$ and the $T_{\mu\nu}$ tensors defined before. We get,
up to first order in $T$:
\begin{eqnarray}
S&=&\int d^4 x \sqrt{\tilde g} i\bar \psi \tilde
e_i^{\;\mu}\gamma^i(\partial_\mu+\tilde\Omega_\mu)\psi+
\frac{1}{2}\int d^4 x \sqrt{\tilde g}iT^\mu_{\;\mu} \bar \psi
\tilde
e_i^{\;\mu}\gamma^i(\partial_\mu+\tilde\Omega_\mu)\psi\nonumber \\
&-&\frac{1}{2}\int d^4 x \sqrt{\tilde g} i\bar \psi
T_i^{\;\mu}\gamma^i(\partial_\mu+\tilde\Omega_\mu)\psi+\frac{1}{4}\int
d^4 x \sqrt{\tilde g} i\bar \psi \tilde
e_i^{\;\mu}\gamma^i(T^\nu_{\;\nu ;\mu}-T^\nu_{\;\mu ;
  \nu})\psi\nonumber\\
\left.\right.
\end{eqnarray}

In the simplest case, in which the Kaluza--Klein gauge bosons are
absent, we can expand the Dirac action in powers of the properly
normalized Goldstone bosons $\pi^\alpha$. Up to $\Od(p^2)$, we
obtain:
\begin{eqnarray}
S&=&\int d^4 x \sqrt{\tilde g} i\bar \psi \tilde
e_i^{\;\mu}\gamma^i(\partial_\mu+\tilde\Omega_\mu)\psi\nonumber
\\&-& \frac{i}{2f^4}\int d^4 x \sqrt{\tilde
g}h_{\alpha\beta}(\pi)\tilde g^{\mu\nu}\partial_\mu
\pi^\alpha\partial_\nu\pi^\beta\bar \psi \tilde
e_i^{\;\rho}\gamma^i(\partial_\rho+\tilde\Omega_\rho)\psi\nonumber
\\ &+&\frac{i}{2f^4}\int d^4 x \sqrt{\tilde g} \bar \psi
h_{\alpha\beta}(\pi)\tilde g^{\mu\nu}\partial_\mu
\pi^\alpha\pabar\pi^\beta(\partial_\nu+\tilde\Omega_\nu)\psi
\nonumber
\\&-&\frac{i}{4f^4}\int d^4 x \sqrt{\tilde g} \bar \psi
h_{\alpha\beta}(\pi)\tilde g^{\mu\nu}(\pabar(\partial_\mu
\pi^\alpha\partial_\nu\pi^\beta) -\partial_\mu(\pabar
\pi^\alpha\partial_\nu\pi^\beta))\psi.
\end{eqnarray}

In particular, this way of introducing the couplings of Goldstone
bosons to fermions allows us to consider chiral fermions in a
straightforward way. Thus for the fermionic sector of the Standard
Model we get:
\begin{eqnarray}
S&=&i\int d^4 x \sqrt{\tilde g} \left(\bar {\cal Q} \Dbar^Q {\cal
Q}+\bar {\cal L} \Dbar^L {\cal L}\right) \nonumber\\
&-&\frac{i}{2f^4}\int d^4 x \sqrt{\tilde
g}h_{\alpha\beta}(\pi)\tilde g^{\mu\nu}\partial_\mu
\pi^\alpha\partial_\nu\pi^\beta\left(\bar {\cal Q} \Dbar^Q {\cal
Q}+\bar {\cal L} \Dbar^L {\cal L}\right)\nonumber
\\ &+&\frac{i}{2f^4}\int d^4 x \sqrt{\tilde g}h_{\alpha\beta}(\pi)
\tilde g^{\mu\nu} \left( \bar {\cal Q}
\partial_\mu \pi^\alpha\pabar\pi^\beta D_\nu^Q {\cal
Q} +\bar {\cal L}\partial_\mu \pi^\alpha\pabar\pi^\beta D_\nu^L
{\cal L}\right)\nonumber
\\&-&\frac{i}{4f^4}\int d^4 x \sqrt{\tilde g}h_{\alpha\beta}(\pi)
\tilde g^{\mu\nu} \left[\bar {\cal Q} (\pabar(\partial_\mu
\pi^\alpha\partial_\nu\pi^\beta) -\partial_\mu(\pabar
\pi^\alpha\partial_\nu\pi^\beta)){\cal Q}\right. \nonumber \\
&+&\left.\bar {\cal L} (\pabar(\partial_\mu
\pi^\alpha\partial_\nu\pi^\beta) -\partial_\mu(\pabar
\pi^\alpha\partial_\nu\pi^\beta)){\cal L}\right],
\end{eqnarray}
where we have:
\begin{eqnarray}
\Dbar^Q&=&\tilde e_i^{\;\mu}\gamma^i D_\mu^Q=\tilde
e_i^{\;\mu}\gamma^i(\partial_\mu+\tilde\Omega_\mu+G_\mu+W_\mu
P_L+ig'(Y_L^Q P_L+Y_R^Q P_R)B_\mu)\nonumber \\ \Dbar^L&=&\tilde
e_i^{\;\mu}\gamma^i D_\mu^L=\tilde
e_i^{\;\mu}\gamma^i(\partial_\mu+\tilde\Omega_\mu+W_\mu
P_L+ig'(Y_L^L P_L+Y_R^L P_R)B_\mu)
\end{eqnarray}
and $Y_{L,R}^{Q,L}$ denote the hypercharge matrices for left or
right quarks and leptons. The left and right projectors $P_{L,R}$
are defined as usual from the four-dimensional $\gamma_5$ matrix.
 The Yukawa sector can be obtained in a straightforward way,
and we obtain:
\begin{eqnarray}
S_{YK}=-\int d^4 x \sqrt{\tilde
g}\left(1-\frac{1}{2f^4}h_{\alpha\beta}(\pi)\tilde
g^{\mu\nu}\partial_\mu \pi^\alpha\partial_\nu \pi^\beta
\right)\left( \bar{\cal Q}_L \Phi H_Q {\cal Q}_R+ \bar{\cal
L}_L\Phi H_L{\cal L} \right) + \mbox{h.c}\nonumber \\
\left.\right.
\end{eqnarray}
Here $H_{Q,L}$ denote the Yukawa matrices and $\Phi$ is the Higgs
doublet.

\subsection*{Gauge bosons}

For the Yang--Mills action on the brane we can follow similar
steps, and we find:
\begin{eqnarray}
S_{YM}&=&\frac{\tr}{2g^2}\int d^4x \sqrt{g}
g^{\mu\rho}g^{\nu\sigma} G_{\mu\nu} G_{\rho\sigma}
=\frac{\tr}{2g^2}\int d^4x \sqrt{\tilde g} \tilde
g^{\mu\rho}\tilde g^{\nu\sigma} G_{\mu\nu} G_{\rho\sigma}\nonumber
\\ &+& \frac{\tr}{4g^2}\int d^4x \sqrt{\tilde g} T^\mu_{\;\mu}\tilde
g^{\mu\rho}\tilde g^{\nu\sigma} G_{\mu\nu} G_{\rho\sigma}
-\frac{\tr}{g^2}\int d^4x \sqrt{\tilde g} T^{\mu\rho}\tilde
g^{\nu\sigma} G_{\mu\nu} G_{\rho\sigma}\nonumber \\ \left.\right.
\end{eqnarray}
where $T_{\mu\nu}$ is defined in (\ref{TT}). Finally expanding the
$T$ terms in Goldstone bosons, we get:
\begin{eqnarray}
S_{YM}&=&\frac{\tr}{2g^2}\int d^4x \sqrt{\tilde g} \tilde
g^{\mu\rho}\tilde g^{\nu\sigma} G_{\mu\nu} G_{\rho\sigma}\nonumber
\nonumber\\&-& \frac{\tr}{4g^2f^4}\int d^4x \sqrt{\tilde g} \tilde
g^{\mu\nu}h_{\alpha\beta}(\pi)(\partial_\mu \pi^\alpha\partial_\nu
\pi^\beta)\tilde g^{\mu\rho}\tilde g^{\nu\sigma} G_{\mu\nu}
G_{\rho\sigma}\nonumber
\\ &+&\frac{\tr}{g^2f^4}
\int d^4x \sqrt{\tilde g}h_{\alpha\beta}(\pi)(\partial_\lambda
\pi^\alpha\partial_\kappa \pi^\beta) \tilde g^{\mu\lambda}\tilde
g^{\rho\kappa}\tilde g^{\nu\sigma} G_{\mu\nu} G_{\rho\sigma}
\end{eqnarray}
From the above discussion we see that the GB always interact by
pairs with the SM matter. In addition, due to their geometric
origin, those interactions are very similar to the gravitational
interactions since the $\pi$ fields couple to all the matter
fields with the same strength, which is suppressed by a factor
$f^4$. This is quite interesting since it could explain why they
have
 not
been observed so far, provided they exist at all. However,
moderate values
 of the brane tension around the TeV scale could make their
 production possible in the next generation of colliders.

\section{Conclusions}

In this work we have studied the effective action describing the
low-energy dynamics of the GB, which appear when the
higher-dimensional space-time manifold isometry group is
spontaneously broken by the presence of a three-brane Universe.
From the $3+1$-dimensional point of view, those GB can be
considered as some kind of new scalar fields whose dynamics is
given by the non-linear sigma model lagrangian corresponding to
the coset manifold $K=G/H$. Eventually, the GB can also get some
mass terms due to small deviations from the simple ideal exact
isometry pattern.

This spontaneous symmetry breaking gives rise, through the Higgs
mechanism, to a mass matrix for the KK graviphotons associated to
the isometries of the compactified space $B$. However, for the
interesting case of large extra dimensions and $f \ll M_D$, the
graviphotons decouple from the low-energy theory and their masses
become very small. We can thus consider the GB as the only
relevant degrees of freedom on the brane in the low-energy regime.

In order to make further studies of the possible phenomenological
implications of those GB brane excitations, we have considered
their corresponding couplings with the SM particles including
scalars, fermions (chiral and non-chiral) and gauge bosons.

In the above-mentioned scenario of large extra dimensions with $f
\ll  M_D$, for appropriate values of $f$ and the GB masses, these
scalar particles are the only brane states that could be probed in
the next generation of colliders such as the LHC. In this case,
the effective couplings obtained here provide the necessary tools
to compute the cross sections and the expected rates of new exotic
events in terms of $f$ and the GB masses only. Work is in progress
in that direction.

 \vspace{.5cm}

 {\bf Acknowledgements:} This work
has been partially supported by the Ministerio de Educaci\'on y
Ciencia (Spain) (CICYT AEN 97-1693 and PB98-0782).    \\

\end{document}